\documentclass[reprint,nofootinbib,amsmath,amssymb,aps]{revtex4-1}
\usepackage{graphicx}
\usepackage{dcolumn}
\usepackage[colorlinks,linkcolor=magenta,anchorcolor=cyan,citecolor=blue]{hyperref}
\usepackage{bm}
\usepackage[multiple]{footmisc}
\newcommand{\fig}[1]{Fig.\ref{#1}}
\def\be{\begin{equation}}
\def\ee{\end{equation}}
\def\ba{\begin{eqnarray}}
\def\ea{\end{eqnarray}}

\def\nn{\nonumber}
\def\lf{\left}
\def\rt{\right}

\newcommand{\eq}[1]{(\ref{#1})}

\def\lf{\left}\def\rt{\right}\def\q{\theta} \def\w{\omega}  \def\y {\psi}   \def\p {\pi}  \def\s {\sigma} \def\d {\delta} \def\f {\phi} \def\g {\gamma}   \def\k {\kappa} \def\l {\lambda} \def\z {\zeta} \def\x {\xi} \def\c {\chi} \def\b {\beta}   \def\pd {\partial}\def\p {\pi} \def \inf {\infty}  
\def\Q{\Theta}      \def\S {\Sigma}  \def\F {\Phi}      \def\grad{\nabla}\def\.{\cdot}
\def\math {\mathcal}
\begin{document}

\title{Weak cosmic censorship conjecture in Einstein-Born-Infeld black holes}
\author{Yan-Li He$^{2}$}
\email{heyanli16@mails.ucas.ac.cn}
\author{Jie Jiang$^{1}$}
\email{Corresponding author. jiejiang@mail.bnu.edu.cn}
\affiliation{$^{1}$Department of Physics, Beijing Normal University, Beijing, 100875, China}
\affiliation{$^{2}$School of Physics, University of Chinese Academy of Sciences, Beijing 100049, China}
\date{\today}

\begin{abstract}
Recently, Sorce and Wald have suggested a new version of the gedanken experiments to overspin or overcharge the Kerr-Newman black holes in Einstein-Maxwell gravity. Following their setup, in this paper, we investigate the weak cosmic censorship conjecture (WCCC) in the static Einstein-Born-Infeld black holes for the Einstein gravity coupled to nonlinear electrodynamics. First of all, we derive the first two order perturbation inequalities of the charged collision matter in the Einstein-Born-Infeld gravity based on the Iyer-Wald formalism as well as the null energy conditions of the matter fields and show that they share the same form as these in Einstein-Maxwell gravity. As a result, we find that the static Einstein-Born-Infeld black holes cannot be overcharged under the second-order approximation after considering these inequalities. Our result at some level hints at the validity of the weak cosmic censorship conjecture for string theory.
\end{abstract}
\maketitle

\section{Introduction}
A naked singularity, which is not hidden behind a black hole horizon will lead to the invalidity of predictability and deterministic nature of general relativity as classical theory. Therefore, Penrose formulated the weak cosmic censorship conjecture (WCCC) to postulate that any physical process cannot destroy the event horizon of the black holes \cite{RPenrose}. To test the validity of the WCCC, in $1974$, Wald proposed a gedanken experiment to overspin or overcharge an extremal Kerr-Newman (KN) black hole by throwing a test partical\cite{Wald94}. He showed that an extremal KN black hole cannot be destroyed in this way. However, it was found by Hubeny\cite{Hubeny,1,2,3,4,5} that the nearly extremal KN black hole can be destroyed by absorbing a test particle, which received lots of attention and was followed by extensive studies in various theories\cite{B1, B2, B3, B4, B5, B6,B7,B8,B9,B10,B11,B12,B13,B14,B15,B16,B17}.

However, there are two important assumptions underlying the above experiments. The collision matter in consideration is a test particle and the correction of the conserved charges for the black hole is only at the level of the first-order perturbation. Therefore, Sorce and Wald \cite{SW} have recently suggested a new version of gedanken experiments where the second-order correction of the mass, angular momentum, and charge are taken into account. In this version, they straightly consider the collision matter source rather than the test particle. Based on the Iyer-Wald formalism \cite{IW} as well as the null energy condition, they obtained the first two perturbation inequalities of the collision matters. As a result, they showed that the WCCC is still valid for the KN black holes under the second-order approximation with consideration of the second-order perturbation inequality.

Most recently, this new version has also been studied in the $5$-dimensional Myers-Perry black holes \cite{An:2017phb}, higher-dimensional charged black holes \cite{Ge:2017vun}, charged dilaton black holes \cite{Jiang:2019ige}, RN-AdS black holes \cite{WJ} as well as Kerr-Sen black holes \cite{Jiang:2019vww}. Although they showed that WCCC is well protected for the nearly extremal case of these black holes after considering the second-order perturbation inequality, there is still a lack of the general proof of the WCCC. Therefore, it is necessary for us to test it in various theories, especially those of some remarkable different properties. We can see that all of the theories mentioned above are only coupled to the linear gauge fields. However, if the quantum corrections or the string modifications are taken into account, nonlinear terms of the gauge fields should be added to the Einstein-Hilbert action. Thus, we want to ask whether the WCCC is also valid for the Einstein gravity with nonlinear electrodynamics which is modified by the string theory. And the investigation of the WCCC in these theories can give a further understanding of the causal structure of the string theory. As one of the most interesting gravitational theories coupled to nonlinear electrodynamics, the Born-Infeld electrodynamics is the effective theory of the vector modes of open string and dynamics of D-branes\cite{Gibbons,deA}. Therefore, in the following of this paper, we would like to consider the Hubeny scenario in the Einstein-Born-Infeld (EBI) black holes by this new version of gedanken experiments and investigate whether the WCCC can be restored when the second-order correction is taken into consideration.

Our paper is organized as follows. In the next section, we review the Iyer-Wald formalism for general diffeomorphism-covariant theories and show the corresponding variational quantities. In Sec. \ref{sec3}, we focus on a static charged black holes and show the corresponding conserved charges in the $4$-dimensional Einstein-Born-Infeld gravity. In Sec. \ref{sec4}, we present the setup for the new version of the gedanken experiment and derive the first two order perturbation inequalities for the optimal first-order perturbation of the EBI black holes. In Sec. \ref{sec5}, we examine the Hubeny scenario from the new version of the gedanken experiment when the second-order perturbation inequality is considered. Finally, the conclusions are presented in Sec. \ref{sec6}.
\\
\\

\section{Iyer-Wald formalism in a diffeomorphism-covariant gravity}\label{sec2}

The new version of the gedanken experiments proposed by Sorce and Wald to destroy a black hole begins with considering an off-shell variation of a general diffeomorphism-covariant theory on a four-dimensional oriented manifold $\math{M}$ with the lagrangian four-form $L$ which is constructed locally out of the metric $g_{ab}$ and other fields $\y$. Following the notation in \cite{IW}, we use $\f=(g_{ab},\y)$ to denote all dynamical fields. Performing a variation of the lagrangian $\bm{L}$, we have
\ba\begin{aligned}\label{varL}
\d \bm{L}=\bm{E}_\f\d\f+d\bm{\Q}(\f,\d\f)\,,
\end{aligned}\ea
where $\bm{E}_\f=0$ gives the equations of motion (EOM) of this gravitational theory, and $\bm{\Q}$ is the symplectic potential three-form which is a linear function of $\d \f$. The symplectic current three-form can be defined by
\ba\begin{aligned}
\bm{\w}(\f,\d_1\f, \d_2\f)=\d_1\bm{\Q}(\f,\d_2\f)-\d_2\bm{\Q}(\f,\d_1\f)\,.
\end{aligned}\ea
If the variation is generated by an infinitesimal diffeomorphism which is related to the vector field $\z^a$, we can replace $\d$ by $\math{L}_\z$ in \eq{varL}. Then, the Noether current three-form $\bm{J}_\z$ associated with $\z^a$ can be defined by
\ba\label{defJ}
\bm{J}_\z=\bm{\Q}(\f, \math{L}_\z\f)-\z\.\bm{L}\,.
\ea
It has been shown in \cite{IW} that this Noether current can be generally expressed by
\ba\label{JQ}
\bm{J}_\z=\bm{C}_\z+d\bm{Q}_\z
\ea
where $\bm{Q}_\z$ is the two-form Noether charge related to $\z^a$ and $\bm{C}_\z=\z^a\bm{C}_a$ are the constraints of the theory, i.e., $\bm{C}_a=0$ for the on-shell dynamical fields.

By virtue of the diffeomprphism invariance, we can keep $\z^a$ fixed. If $\z^a$ is a Killing vector field and the background fields satisfy the EOM, we can further obtain the first two variation identities of the collision matter,
\ba\begin{aligned}\label{var1}
d[\d\bm{Q}_\z-\z\.\bm{\Q}(\f,\d\f)]&=\bm{\w}\lf(\f,\d\f,\math{L}_\z\f\rt)-\z\.\bm{E}\d\f-\d \bm{C}_\z,\\
d[\d^2\bm{Q}_\z-\z\.\d\bm{\Q}(\f,\d\f)]&=\bm{\w}\lf(\f,\d\f,\math{L}_\z\d\f\rt)\\
&-\z\.\d\bm{E}\d\f-\d^2 \bm{C}_\z\,.
\end{aligned}\ea

As mentioned in the introduction, in what follows, we would like to study the static EBI black holes. Therefore, here we consider the case where the background spacetime is asymptotic flat and static with a timelike Killing vector field $\x^a$ which is normalized at asymptotic infinity. Utilizing this Killing vector, the ADM mass of this black hole can be expressed in the following form
\ba\begin{aligned}
\d M&=\int_\inf \lf[\d\bm{Q}_\x-\x\.\bm{\Q}(\f,\d\f)\rt]\,.
\end{aligned}\ea

After replacing $\z$ by $\x$ and integrating the perturbation identities \eq{var1} on the hypersurface $\S$ with a cross section $B$ of the horizon and the spacial infinity as its boundaries, we have
\ba\begin{aligned}\label{var12}
\d M&=\int_{B}\lf[\d\bm{Q}_\x-\x\.\bm{Q}(\f,\d\f)\rt]-\int_\S \d \bm{C}_\x\,,\\
\d^2 M&=\int_{B}\lf[\d^2\bm{Q}_\x-\x\.\d \bm{Q}(\f,\d\f)\rt]\\
&-\int_\S\x\.\d \bm{E}\d\f-\int_\S \d \bm{C}_\x+\math{E}_\S(\f,\d\f)\,,
\end{aligned}\ea
where we denote
\ba
\math{E}_\S(\f,\d\f)=\int_\S\bm{\w}(\f,\d\f,\math{L}_\x\d\f)\,.
\ea

\section{EBI gravity and its static solution}\label{sec3}
In this section, we consider the four-dimensional EBI theory \cite{BI,Hoffmann}, where the lagrangian can be expressed by the following form
\ba
\bm{L}=\frac{1}{16\p}\lf[R-h(\math{F})\rt]\bm{\epsilon}
\ea
with the nonlinear electromagnetic term
\ba
h(\math{F})=\frac{\b^2}{4}\lf(1-\sqrt{1+\frac{\math{F}}{2\b^2}}\rt)\,,
\ea
where we denote $\math{F}=F_{ab}F^{ab}$ with the electromagnetic strength $\bm{F}=d\bm{A}$, $\b$ is the Born-Infeld parameter being equal to the maximum value of electromagnetic field intensity.  According to this action, we can obtain the symplectic potential,
\ba\begin{aligned}
\bm{\Q}(\f,\d\f)=\bm{\Q}^\text{GR}(\f,\d\f)+\bm{\Q}^{\text{BI}}(\f,\d\f)
\end{aligned}\ea
with
\ba\begin{aligned}\label{qqq}
\bm{\Q}_{abc}^\text{GR}(\f,\d\f)&=\frac{1}{16\p}\epsilon_{dabc}g^{de}g^{fg}\lf(\grad_g \d g_{ef}-\grad_e\d g_{fg}\rt)\,,\\
\bm{\Q}_{abc}^\text{BI}(\f,\d\f)&=-\frac{1}{4\p}\epsilon_{dabc}G^{de}\d A_e\,.
\end{aligned}\ea
Here we have defined
\ba
\bm{G}=h'(\math{F})\bm{F}\,.
\ea
The Noether charge is given by
\ba
\bm{Q}_\x=\bm{Q}_\x^\text{GR}+\bm{Q}_\x^\text{BI}\,,
\ea
where
\ba\begin{aligned}
\lf(\bm{Q}_\x^\text{GR}\rt)_{ab}&=-\frac{1}{16\p}\epsilon_{abcd}\grad^c\x^d\,,\\
\lf(\bm{Q}_\x^\text{BI}\rt)_{ab}&=-\frac{1}{8\p}\epsilon_{abcd}G^{cd}A_e\x^e\,.\\
\end{aligned}\ea
The constraints can be shown as
\ba\begin{aligned}\label{EC}
\bm{C}_{abcd}&=\epsilon_{ebcd}\lf(T_a{}^e+A_aj^e\rt)\,,
\end{aligned}\ea
where we denote
\ba\begin{aligned}\label{TJ}
T_{ab}=\frac{1}{8\p}G_{ab}-T_{ab}^\text{BI}\,,\ \ \
j^a=\frac{1}{4\p}\grad_a G^{ab}\,,
\end{aligned}\ea
with
\ba\begin{aligned}\label{TTT}
T_{ab}^\text{BI}&=\frac{1}{4\p}\lf[G_{ac}F_b{}^c-\frac{1}{4}g_{ab}h(\math{F})\rt]\,.\\
\end{aligned}\ea
If there are some additional charged matter sources, $T^{ab}$ and $j^a$ are nonvanishing and they correspond to the stress-energy tensor as well as the electromagnetic charge-current of the additional matter. And $T^{ab}=j^a=0$ gives the EOM of the on-shell fields. As mentioned above, here the background spacetime we considered is static, which means $\math{L}_\x \bm{A}=\x\. \bm{F}+d\lf(\x\.\bm{A}\rt)=0$. Since $\x\. \bm{A}$ is a constant on the horizon, we have
\ba\label{Fform}
\x_{a}G^{ab}\propto\x_{a}F^{ab}\propto \x^b
\ea
on the horizon. From \eq{qqq}, the symplectic current for the EBI theory can be written as
\ba
\bm{\w}(\f, \d_1\f,\d_2\f)=\bm{\w}_{abc}^\text{GR}+\bm{\w}_{abc}^\text{BI}\,,
\ea
where
\ba\begin{aligned}\label{3w}
\bm{\w}_{abc}^\text{GR}&=\frac{1}{16\p}\epsilon_{dabc}w^d\,,\\
\bm{\w}_{abc}^\text{BI}&=\frac{1}{4\p}\lf[\d_2\lf(\epsilon_{dabc}G^{de}\rt)\d_1 A_e -\d_1\lf(\epsilon_{dabc}G^{de}\rt)\d_2 A_e \rt]\,,\\
\end{aligned}\ea
in which we denote
\ba\begin{aligned}
w^a=P^{abcdef}\lf(\d_2g_{bc}\grad_d\d_1 g_{ef}-\d_1 g_{bc}\grad_d\d_2g_{ef}\rt)
\end{aligned}\ea
with
\ba\begin{aligned}
P^{abcdef}&=g^{ae}g^{fb}g^{cd}-\frac{1}{2}g^{ad}g^{be}g^{fc}-\frac{1}{2}g^{ab}g^{cd}g^{ef}\\
&-\frac{1}{2}g^{bc}g^{ae}g^{fd}+\frac{1}{2}g^{bc}g^{ad}g^{ef}\,.
\end{aligned}\ea
\begin{figure}
\begin{center}
\includegraphics[width=0.49\textwidth,height=0.28\textheight]{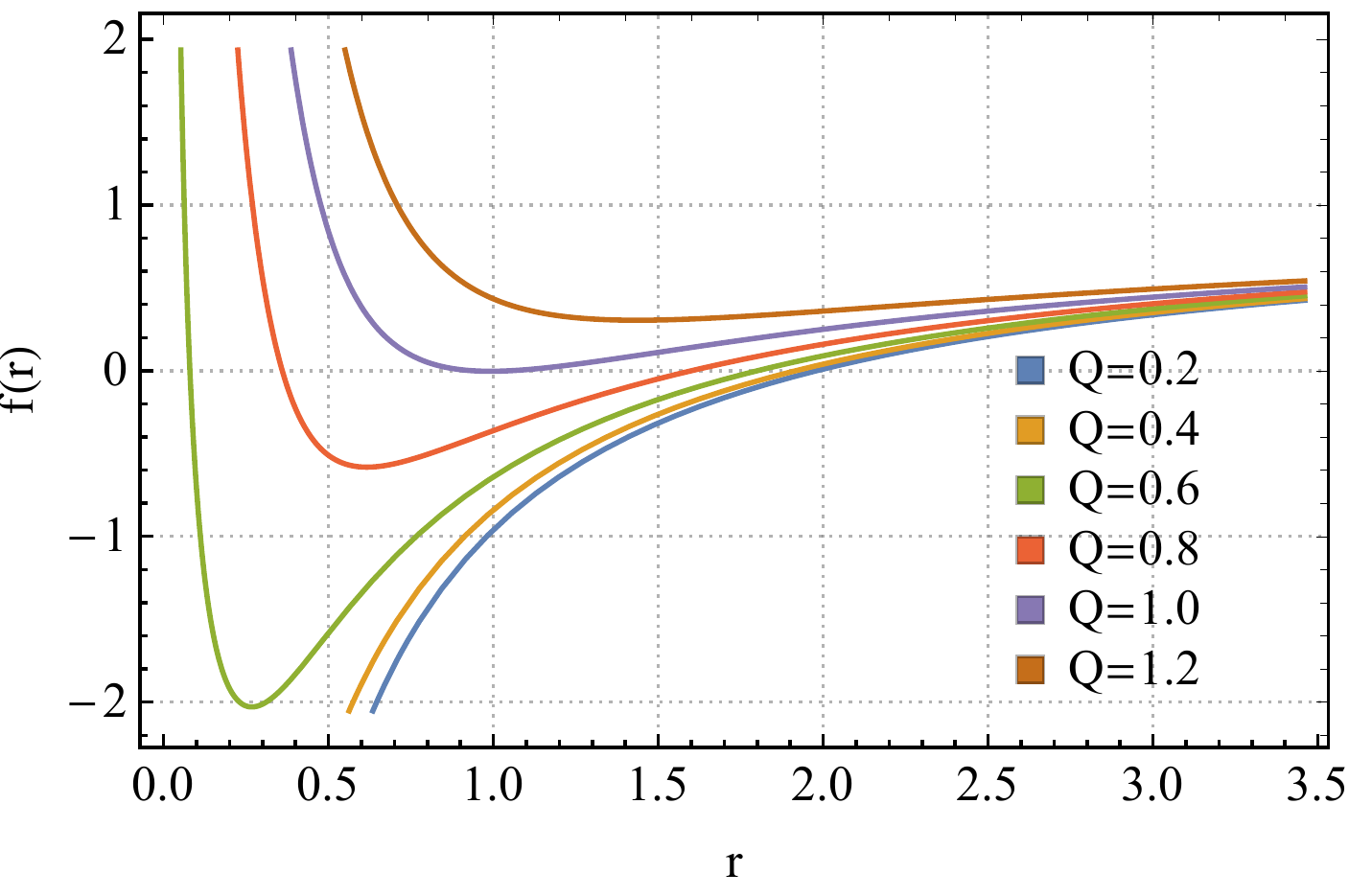}
\caption{Plot showing the radial dependence of blackening factor $f(r)$ for different values of electric charge $Q$. Here we have set $M=1$ and $\b=4$.}\label{fig}
\end{center}
\end{figure}

We next restrict on the static EBI black hole solution in this theory. The line element as well as the electromagnetic field can be read off \cite{Hoffmann}
\ba\begin{aligned}\label{ds2}
ds^2&=-f(r)dt^2+\frac{1}{f(r)}dr^2+r^2d\q^2+r^2\sin^2\q d\f^2\,,\\
\bm{A}&=-\frac{Q \math{K}(r)}{r} dt\,,
\end{aligned}\ea
where
\ba\begin{aligned}
f(r)=1-\frac{2M}{r}+\frac{q(r)^2}{r^2}
\end{aligned}\ea
is the blackening factor with
\ba\begin{aligned}
q^2(r)&=\frac{2\b^2r^4}{3}\lf(1-\sqrt{1+\frac{Q^2}{\b^2 r^4}}\rt)+\frac{4Q^2\math{K}(r)}{3}\,.
\end{aligned}\ea
Here we have denoted
\ba
\math{K}(r)={}_2F_1\lf(\frac{1}{4},\frac{1}{2},\frac{5}{4},-\frac{Q^2}{\b^2r^4}\rt)
\ea
and ${}_2F_1$ is the Gaussian hypergeometric function, $M$ and $Q$ are associated with the mass and electric charge of this black hole. According to \fig{fig}, we can see that this line element \eq{ds2} has two types of black hole solutions, i.e., the single-horizon and double-horizons black hole solutions. The radius $r_h$ of the event horizon is the largest root of $f(r_h)=0$. The corresponding surface gravity, area, and electric potential of the black hole can be shown as
\ba\begin{aligned}\label{kAWF}
\k&=\frac{f'(r_h)}{2}\,,\ \ A_\math{H}=4\p r_h^2\,,\ \ \F_\math{H}=\frac{Q\math{K}(r_h)}{r_h}\,.
\end{aligned}\ea

For the double-horizons case, there exists an extremal black hole solution. By solving the equations $f'(r_e)=f(r_e)=0$, we can obtain
\ba\begin{aligned}\label{excond}
M&=\frac{r_e}{3}+\lf(\frac{2r_e}{3}+\frac{1}{6r_e\b^2}\rt)\math{K}(r_e)
\,,\\
Q^2&=r_e^2+\frac{1}{4\b^2}
\end{aligned}\ea
with the horizon radius $r_e$ for the extremal EBI black hole.

\section{Perturbation inequalities in EBI gravity}\label{sec4}
In this section, following the setup of the new version of gedanken experiments proposed by Sorce and Wald \cite{SW}, we turn to derive the second two order inequalities in the EBI gravitational theory based on the null energy condition of the matter fields. Now, we consider the situation where the static EBI black hole is perturbed by a one-parameter family additional charged matter source which is only nonvanishing in a compact region of the future horizon. The EOM of the dynamical fields can be written as
\ba\begin{aligned}
&R_{ab}(\l)-\frac{1}{2}R(\l)g_{ab}(\l)=8\p \lf[T_{ab}^\text{BI}(\l)+T_{ab}(\l)\rt]\,,\\
&\grad_a^{(\l)}G^{ab}(\l)=4\p j^a(\l)\,.
\end{aligned}\ea
Because the background geometry is the EBI black hole, we have $T^{ab}=0$ and $j^a=0$ for the background fields, where we denote $\c=\c(0)$ to the background quantity $\c$. With a similar consideration to \cite{SW}, here the EBI black holes are also assumed to be linearly stable under perturbations, i.e., any source-free solution to the linearized equation of motion \eq{ds2} will approach a perturbation towards another EBI black hole at sufficiently late times. Therefore, we can always choose the hypersurface such that $\S=\math{H}\cup \S_1$. Here $\math{H}$ is a portion of the future horizon which is bounded by the bifurcate surface $B$ of the background spacetime as well as the very late cross section $B_1$ where the matter source vanishes. $\S_1$ is a spacelike hypersurface connected $B_1$ and spatial infinity where the dynamical fields is described by the static EBI solutions \eq{ds2}.

According to above setup, we can see that the perturbation vanishes on the bifurcation surface $B$. Since the background fields are source free and satisfy the EOM $\bm{E}_\f=0$, the first-order perturbation identity \eq{var12} can be reduced to
\ba\begin{aligned}\label{fst1}
\d M&=-\int_\math{H}\epsilon_{ebcd}\lf[\d T_a{}^e+A_a\d j^e\rt]\x^a\,,
\end{aligned}\ea
where we used the fact that $T^{ab}=j^a=0$ in the background spacetime. Without loss of generalization, we can choose the gauge of the electromagnetic field $\bm{A}$ such that the electric potential $\F=-\x^a A_a$ vanishes at asymptotic infinity. Then, it will be a constant on $\math{H}$. According to the electromagnetic part of \eq{TJ}, we can further obtain
\ba\begin{aligned}
&-\int_\math{H}\epsilon_{ebcd}\x^aA_a\d j^e=\F_\math{H}\d\lf[\int_\math{H}\epsilon_{ebcd} j^e\rt]\\
&=\frac{1}{4\p}\F_\math{H}\d\lf[\int_\math{H}\epsilon_{ebcd} \grad_a G^{ea}\rt]\\
&=\frac{1}{8\p}\F_\math{H}\d\lf[\int_{B_1}\epsilon_{ebcd} G^{eb}-\int_{B}\epsilon_{ebcd} G^{eb}\rt]\\
&=\F_\math{H}\d Q
\end{aligned}\ea
with the electric charge of the black hole
\ba
Q=\frac{1}{8\p}\int_\inf\epsilon_{ebcd} G^{eb}=\frac{1}{8\p}\int_{B_1}\epsilon_{ebcd} G^{eb}\,.
\ea
Here we have used the Gaussian theorem as well as the facts that the background field is source free and the current $j^a$ vanishes on $\S_1$. Using this result, Eq. \eq{fst1} becomes
\ba\label{var1eq1}\begin{aligned}
\d M-\F_\math{H}\d Q=\int_\math{H}\bm{\tilde{\epsilon}} \d T_{ab}k^a \x^b\geq 0
\end{aligned}\ea
where the null energy condition $\d T_{ab}k^a k^b\geq 0$ has been used, and $\bm{\tilde{\epsilon}}$ is the volume element on the horizon, which can be obtained from $\epsilon_{ebcd}=-4k_{[e}\tilde{\epsilon}_{bcd]}$ with the future-directed normal vector $k^a \propto \x^a$ on the horizon. From \fig{fig}, we can see that for the single-horizon black holes in the EBI theory, the horizons will not be broken, while for the double-horizons black holes, the naked singularity can be obtained by adding the charge of the spacetime, which means the optimal choice to violate the EBI black hole is to saturate \eq{var1eq1}, i.e., the energy flux through the horizon vanished for the first-order non-electromagnetic perturbation. Then, \eq{var1eq1} comes
\ba\label{var1eq1op}\begin{aligned}
\d M-\F_\math{H}\d Q=0\,.
\end{aligned}\ea

Next, we consider the second-order perturbation inequality under the above optimal condition. By performing a similar analysis to the first-order result, we can further obtain
\ba\begin{aligned}\label{sec22}
\d^2 M&=-\int_\math{H}\x\.\d \bm{E}_\f\d\f-\int_\math{H} \d \bm{C}_\x+\math{E}_\S(\f,\d\f)\\
&=-\int_\math{H} \d \bm{C}_\x+\math{E}_\S(\f,\d\f)\,.
\end{aligned}\ea
Here the integrals in the last two terms only depend on the surface $\math{H}$ because the dynamical fields satisfy the source-free EOM on the hypersurface $\S_1$, i.e., we have $\bm{E}_\f(\l)=\bm{C}(\l)=0$ on $\S_1$. At the last step, we used the fact that $\x^a$ is tangent to the horizon. By considering the optimal condition of the first-order perturbation, i.e., $\d T_{ab}$ vanishes on the horizon, with similar analysis with the first-order perturbation inequality, Eq. \eq{sec22} reduces to
\ba\begin{aligned}\label{dM22}
\d^2M-\F_H\d^2Q&=\math{E}_\S(\f,\d \f)+\int_\math{H}\bm{\tilde{\epsilon}}\d^2T_{ab}\x^ak^b\\
&\geq\math{E}_\math{H}(\f,\d \f)+\math{E}_{\S_1}(\f,\d \f)\,.
\end{aligned}\ea
where we have used the energy condition for the second-order perturbed stress-energy tensor of the collision matter in the last step and imposed the condition $\x^a \d A_a|_\math{H}=0$ by a gauge transformation \cite{SW}. The first term of the right side in \eq{dM22} can be decomposed into
\ba\label{EEE3}
\math{E}_\math{H}(\f,\d \f)=\int_\math{H}\bm{\w}^\text{GR}+\int_\math{H}\bm{\w}^\text{BI}\,.
\ea
According to \cite{SW}, the gravitational part in above expression  is given by
\ba
\int_\math{H}\bm{\w}^\text{GR}=\frac{1}{4\p}\int_{\math{H}}(\x^a\grad_a u)\d\s_{ac}\d\s^{bc}\bm{\tilde{\epsilon}}\geq 0\,.
\ea
For the EBI part, according \eq{3w}, we have
\ba\begin{aligned}\label{EEM}
\bm{\w}^\text{BI}_{abc}&=\frac{1}{4\p}\epsilon_{dabc}\lf[\d A_e\math{L}_\x\d G^{de}-\d G^{de}\math{L}_\x\d A_e\rt]\\
&+\frac{1}{4\p}\lf[(\math{L}_\x\d\epsilon_{dabc})G^{de}\d A_e-\d\epsilon_{dabc}G^{de}\math{L}_\x\d A_e\rt]\,.
\end{aligned}\ea
By considering the gauge condition $\x^a\d A_a=0$ on the horizon as well as the assumption that the background fields satisfy Eq. \eq{Fform}, the last two terms vanish. Then, Eq. \eq{EEM} can be written as
\ba\begin{aligned}\label{EMw}
\bm{\w}^\text{BI}_{abc}&=\frac{1}{4\p}\math{L}_\x\lf(\epsilon_{dabc}\d A_e\d G^{de}\rt)-\frac{1}{2\p}\epsilon_{dabc}\d G^{de}\math{L}_\x\d A_e\,.\\
\end{aligned}\nn\\\ea
Using the Stoke's theorem, the integral over $\math{H}$ of the first term on the right side will only contribute a boundary term at $B_1$. According to the stability assumption, $\f(\l)$ on $B_1$ is the EBI solution. That is to say, $\d G_{ab}$ also satisfies Eq. \eq{Fform}. Together with the gauge condition $\x^a\d A_a=0$ on $\math{H}$, the first term of \eq{EMw} makes no contribution to \eq{EEM}. Combining above results, we have
\ba\begin{aligned}\label{EMw1}
\math{E}_\math{H}(\f,\d\f)&=-\frac{1}{2\p}\int_\math{H}\epsilon_{dabc}\d G^{de}\math{L}_\x\d A_e\\
&=\int_\math{H}\bm{\tilde{\epsilon}}\x^ak^b\lf(\d^2T_{ab}^\text{BI}\rt)\geq 0 \,,\\
\end{aligned}\ea
where we also used the null energy condition for the electromagnetic stress-energy tensor. Finally, \eq{dM22} reduces to
\ba
\d^2M-\F_\math{H}\d^2Q\geq \math{E}_{\S_1}(\f,\d\f)\,.
\ea

Now we are left out to evaluate $\math{E}_{\S_1}(\f,\d\f)$. To calculate it, we follow the trick introduced in \cite{SW}, and write $\math{E}_{\S_1}(\f,\d\f)=\math{E}_{\S_1}(\f,\d\f^\text{BI})$, where $\f^\text{BI}$ is introduced by the variation of a family of dilaton black hole solutions \eq{ds2},
\ba\label{varMQ}\begin{aligned}
M^\text{BI}(\l)&=M+\l \d M\,,\\
Q^\text{BI}(\l)&=Q+\l \d Q\,,
\end{aligned}\ea
where $\d M$ and $\d Q$ chosen to be in agreement with the first order variation of the above optimal perturbation by the matter source. From the variation \eq{varMQ}, one can find $\d^2 M=\d^2 Q=\d \bm{E}=\d^2 \bm{C}=\math{E}_\math{H}(\f,\d \f^\text{BI})=0$. Thus, from the second expression of \eq{var12}, we have
\ba
\math{E}_{\S_1}(\f, \d \f^\text{BI})=-\int_B\lf[\d^2\bm{Q}_\x-\x\.\d \bm{\Q}(\f,\f^\text{BI})\rt]\,.
\ea
Since $\x^a=0$ on the bifurcation surface $B$, we have
\ba
\math{E}_{\S_1}(\f, \d \f^\text{BI})=-\frac{\k}{8\p}\d^2A_B^\text{BI}\,.
\ea
Therefore, the second-order inequality becomes
\ba\label{secorder1}
\d^2M-\F_\math{H}\d^2 Q\geq -\frac{\k}{8\p}\d^2A_B^\text{BI}\,.
\ea
Here $A_B^\text{BI}(\l)$ is the area of the bifurcation surface $B$ for the static EBI black hole with mass $M^\text{BI}(\l)$ and charge $Q^\text{BI}(\l)$. From the line element \eq{ds2} of EBI black holes, we can see that the right sight of the inequality \eq{secorder1} can be evaluated by taking two variations of the area formula $A_B=4\p r_h^2$. Using the fact that
\ba\label{exp22}
f\lf(r_h^\text{BI}(\l), M^\text{BI}(\l), Q^\text{BI}(\l)\rt)=0\,,
\ea
and taking a variation of this equation, we can obtain
\ba\begin{aligned}
\d r_h^\text{BI}=\frac{2r_h \d M-2 Q \math{K}(r_h)}{r_h(1+2r_h^2\b^2-2\b \g)}\,,
\end{aligned}\ea
where we denote
\ba
\g=\sqrt{Q^2+r_h^4 \b^2}\,.
\ea
Considering the optimal choice of the first-order perturbation, we can see that the first-order variation of the horizon radius $\d r_h^\text{BI}$ vanishes. By taking two variation of equation \eq{exp22} and using the optimal condition $\d r_h^\text{BI}=0$,
we can further obtain
\ba\begin{aligned}
\d^2 r_h^\text{BI}=-\frac{\d Q^2[r_h^2\b/\g+\math{K}(r_h)]}{r_h(1+2r_h^2\b^2-\b \g)}\,.
\end{aligned}\ea
Then, we have
\ba\begin{aligned}
\d^2A_B^\text{BI}&=8\p \lf[(\d r_h^\text{BI})^2+r_h \d^2 r_h^\text{BI}\right]\\
&=8\p r_h \d^2 r_h^\text{BI}\\
&=-\frac{8\p\d Q^2[r_h^2\b/\g+\math{K}(r_h)]}{1+2r_h^2\b^2-\b \g}
\end{aligned}\ea
By using the expression of the surface gravity
\ba\begin{aligned}
\k=\frac{1+2r_h^2\b^2-\b \g}{2r_h}\,,
\end{aligned}\ea
the second-order inequality can be further obtained,
\ba\label{secineq}
\d^2M-\F_\math{H}\d^2 Q\geq \frac{\d Q^2\left[r_h^2\b+\g\math{K}(r_h)\right]}{2 \g r_h}.
\ea

\section{Gedanken experiments to destroy a nearly extremal black hole}\label{sec5}
In the last section, we obtained the first two order perturbation inequalities. Next, we will perform them into testing the WCCC in the EBI black holes under the second-order approximation. As mentioned in the last section, we know that the WCCC is valid for the single horizon EBI black holes and therefore we only discuss the double-horizons cases in the following. For a double-horizons black hole, there is not a simple formula like the KN black holes for overcharging condition. However, from \fig{fig}, we can see that there is a negative minimum value for $f(r)$ for any black hole solutions. If this minimal value becomes positive, the black hole horizon is destroyed. Therefore, here we define a function of $\l$ as
\ba\label{cd1}
h(\l)=f\lf(r_m(\l),M(\l),Q(\l)\rt)\,,
\ea
where $r_m(\l)$ is defined as the minimal radius of the function $f(r, M(\l), Q(\l))$, and it can be determined by
\ba\begin{aligned}\label{rmeq}
\pd_rf\lf(r_m(\l),M(\l),Q(\l)\rt)=0\,,
\end{aligned}\ea
which gives
\ba\begin{aligned}
M=\frac{2r_m\b\lf(\g_m-r_m^2\b\rt)}{3}+\frac{2Q^2 \math{K}(r_m)}{3r_m}\,,
\end{aligned}\ea
where we denote
\ba
\g_m=\sqrt{Q^2+r_m^4\b^2}\,.
\ea

According to the stability assumption, the perturbed geometry at the late times should be described by the EBI solution. That is to say, if there exists a spacetime solution $\f(\l)$ such that $h(\l)>0$, the WCCC is violated. Considering the second-order approximation of $\l$, we have
\ba\begin{aligned}
&h(\l)=1+2r_m^2\b^2-2\b \g_m-\frac{2\l}{r_m}\lf(\d M-\frac{Q\math{K}(r_m)\d Q}{r_m}\rt)\\
&+\l^2\lf[\frac{2\b^2\d r_m^2(\g_m-r_m^2\b)}{\g_m}+\frac{\d Q^2}{2}\lf(\frac{\math{K}(r_m)}{r_m^2}+\frac{\b}{\g_m}\rt)\rt]\\
&-\frac{2\l^2 Q \d Q \d r_m\lf(r_m^2\b+\math{K}(r_m)\g_m\rt)}{r_m^3\g_m}+\frac{2\l^2\d M \d r_m}{r_m^2}\\
&-\frac{\l^2}{r_m}\lf(\d^2M-\frac{Q \math{K}(r_m)\d ^2 Q}{r_m}\rt)\,.
\end{aligned}\ea
Taking the variation of \eq{rmeq}, we can obtain
\ba\begin{aligned}
\d r_m=\frac{Q \d Q\lf[r_m^2\b+\math{K}(r_m)\g_m\rt]-r_m \g_m \d M}{2r_m^3\b^2(\g_m-r_m^2\b)}\,.
\end{aligned}\ea

Following a similar setup with \cite{SW}, here we consider the case where the background geometry is a nearly extremal EBI black hole which satisfies $r_m=(1-\epsilon)r_h$ with some small parameter $\epsilon$ chosen to agree with the first-order perturbation of the matter source. Then, Eq. \eq{rmeq} implies that
\ba
\pd_rf\lf(r_h,M,Q\rt)=\epsilon r_h \pd_r^2f\lf(r_h,M,Q\rt)
\ea
under the first-order approximation of $\epsilon$. Therefore, we have
\ba\begin{aligned}
&f(r_m,M,Q)=f\lf(r_h(1-\epsilon), M, Q\rt)\\
&\simeq -\epsilon r_h \pd_r f(r_h,M,Q)+\frac{\epsilon^2 r_h^2}{2} \pd_r^2 f(r_h,M,Q)\\
&=-\frac{1}{2}\epsilon^2 r_h^2 \pd_r^2 f(r_h,M,Q)
\end{aligned}\ea
under the second-order approximation of $\epsilon$, which gives
\ba\begin{aligned}
1+2r_m^2\b^2-2\b \g_m=\lf(1-\frac{2Q^2\b}{\g}\rt)\epsilon^2
\end{aligned}\ea

Utilizing the optimal condition \eq{var1eq1op} of the first-order perturbation as well as the second-order inequality \eq{secineq}, together with above results, we can further obtain
\ba\begin{aligned}
h(\l)&=\lf(1-\frac{2Q^2\b}{\g}\rt)\epsilon^2+\frac{2\l Q\b \d Q \epsilon}{\g}\\
&-\frac{\l^2\d Q^2}{2}\lf(\frac{1}{r_h^2}+\frac{\b}{\g}\rt)\\
&=-\frac{2\b^2(Q\d Q-r_h^2\epsilon)^2}{r^2(1+2\b^2)}
\end{aligned}\ea
under the second-order approximation of the collision matter source. The last step we have replaced $r_h$ by the radius $r_e$ of the extremal case under the second-order approximation since here we can neglect the difference of $r_m$ and $r_h$. The above expression gives $h(\l)<0$ under the second-order perturbation, which implies that the EBI black holes cannot be overcharged when the second-order perturbation is taken into account.

\section{Conclusion}\label{sec6}

Recently, Sorce and Wald suggested a new version of gedanken experiments to test the WCCC and found that it is valid for KN black hole under the second-order approximation of the collision matter source. Following this setup, it also has been investigated in various theories \cite{Jiang:2019ige,Jiang:2019vww,Ge:2017vun,An:2017phb}. However, all of these theories only considered the case of the gravity coupled to nonlinear electrodynamics. After considering the correction of string theory, nonlinear electrodynamics shall be added to the Einstein gravity. Therefore, in this paper, we considered the EBI gravity and tested the WCCC for its static black hole solutions. First, based on the Iyer-Wald formalism as well as the null energy condition of the matter fields, we derived the first two perturbation inequalities in EBI gravity. As a result, we found that the nearly extremal static EBI black holes cannot be overcharged under the second-order approximation. Therefore, there is no violation of the WCCC occurs around the static black holes in EBI gravity. This result might indicate that once this black hole is formed, it will never be overcharged classically. Moreover, our result at some level hints at the validity of the weak cosmic censorship conjecture for string theory.

\section*{acknowledgements}
This research was supported by National Natural Science Foundation of China (NSFC) with Grants No. 11675015.
\\
\\

\end{document}